\documentclass{proclund}

\begin{document}
\pagestyle{prochead}


\title{RECENT DEVELOPMENTS IN RELATIVISTIC MODELS FOR EXCLUSIVE $A(e,e'p)B$ REACTIONS}
\author{J.M. Ud\'{\i}as}
  \email{jose@nuc2.fis.ucm.es}
  \homepage{http://nuclear.fis.ucm.es}
\author{Javier R. Vignote}
\affiliation
 {Dpto. F\'{\i}sica At\'omica, Molecular y Nuclear,  Universidad Complutense de
Madrid,
Spain}
\author{E. Moya de Guerra}
\author{A. Escuderos}
\affiliation{
Instituto de Estructura de la Materia (CSIC), Madrid, Spain}
\author{J.A. Caballero}
\affiliation{ 
Dpto. F\'{\i}sica At\'omica, Molecula y Nuclear, Universidad de Sevilla, Spain\\~\\
}


\begin{abstract}
A comparison of impulse approximation calculations for the $(e,e'p)$ reaction,
based on the Dirac equation and the Schr\"odinger one is presented. Trivial
(kinematics) differences are indicated, as well as how to remove them from the
standard  nonrelativistic formalism.
Signatures of the relativistic approach are found where  the enhancement of the
lower components (spinor distortion or negative energy contributions)  modifies
$TL$ observables with respect to the nonrelativistic predictions, what 
  seems to be confirmed
by the experiment. Finally, the relativistic approach  is used to analyze several 
experiments 
for the reaction $^{16}O(e,e'p)^{15}N$ taken at values of  $Q^2$  from $0.2$ to $0.8$ $(GeV/c)^2$, not
finding a significant  $Q^2$ dependence of the scale factors over this range.
\end{abstract}
\maketitle
\setcounter{page}{1}

\section{Introduction}

Quasielastic $(e,e'p)$ processes are a powerful tool to study bound
 nucleon properties.
 Indeed, coincidence $(e,e'p)$ measurements at quasielastic kinematics
 have provided over the years detailed information on the energies,
 momentum distributions and spectroscopic factors of bound nucleons.
 This is so because at quasielastic kinematics the $(e,e'p)$ reaction
 can be treated with confidence in the impulse approximation, {\em i.e.},
 assuming that the detected  proton absorbs the whole
 momentum ($q$) and energy ($\omega$) of the exchanged photon (for recent
 reviews of the subject see ref.~\cite{BGPR96} and references
  therein). Until
 recently, most data were concentrated in the low missing momentum range
 $p_m\leq 300$ MeV/c, where $p_m$ is the recoil momentum of the
  residual nucleus. In the last years~\cite{Gao2000} higher $p_m$-regions
   are
 being probed at small missing energies ($E_m$) to study
 further aspects of bound nucleon dynamics and nucleon currents.                                              

The higher momentum transfer employed in the new experiments, made almost
compulsory a fully relativistic treatment of the reaction mechanism. Within the 
most simple relativistic framework, that parallells  usual nonrelativistic
approaches, 
a single-particle equation is used to compute both the bound and ejected nucleon wave function.
The difference between the nonrelativistic approach employed in the 70's and 80's and the
relativistic formalism lies thus, mainly, in the use of the Dirac equation instead
of the Schr\"odinger one. The success of the simple Dirac single-particle picture 
in describing detailed features of the $(e,e'p)$ experimental data is intringuing and
deserves close examination. In this contribution we compare in detail the 
nonrelativistic and relativistic impulse approximation  to
$(e,e'p)$.

\section{Formalism}

In the relativistic distorted wave impulse
approximation (RDWIA)\cite{Udi93,Udi95,Udi96,Udi99,vanorden,Jin94,Gardner}
 the one-body nucleon current, written for convenience in
$p-$space,
\begin{equation}
J^{\mu}_{N}(\omega,\vec{q})=\int\/\/ d\vec{p}\/
\bar{\psi}_F
(\vec{p}+\vec{q}) \hat{J}^\mu_N(\omega,\vec{q}\/) \psi_B(\vec{p})\; ,
\label{nucc}
\end{equation}
is calculated with relativistic   $\psi_B$ and $\psi_F$
 wave functions for  initial bound
and  final outgoing  nucleons, respectively, and with relativistic nucleon
current operator, $\hat{J}^\mu_N$.
The bound state wave function is a four-spinor with well-defined parity and
angular momentum quantum numbers, and is
obtained by solving the Dirac equation with scalar-vector
(S-V) potentials determined through a Hartree procedure from a relativistic
Lagrangian with scalar and vector meson terms~\cite{HMS91}.
 The wave function for the outgoing proton is a  solution
 of the Dirac equation containing  S-V global optical potentials~\cite{CHCM93}
 for a nucleon scattered with asymptotic momentum $\vec{P}_F$.
Dirac equations for both scattered and bound wave functions are solved
in coordinate space and their solutions are then transformed to
momentum space where necessary.
                                                                                
Eq.~(1) sets up the scenario where differences between 
 the  relativistic and nonrelativistic impulse approximation
approaches are at play.
 As a guide in the comparison with the nonrelativistic formalism
we  study the so-called  {\em factorized approximation}, not because it is often used
(nowadays the simple factorized results are hardly employed to analyze exclusive
$(e,e'p)$ experiments) but because of its simplicity and usefulness as a pedagogical
tool.

\subsection{Factorization}

The $(e,e'p)$ reaction has a simple interpretation if one assumes 
a factorized expression for the cross-section, {\em i.e.}: 

\begin{equation}
\frac{d^5\sigma}{d\Omega_ed\varepsilon'd\Omega_F}=  K \sigma_{ep} S(E_m,\vec{p}_m)
\label{fac}
\end{equation}
where $K$ is the phase-space factor, $\sigma_{ep}$ is the electron-proton
cross-section and $S(E_m,\vec{p}_m)$ is the {\em spectral function} representing the
probability of finding a proton in the target nucleus with missing energy $E_m$ and
missing momentum $p_m$, compatible with the kinematics of the reaction. 
Distortion of the electron wave functions due to the Coulomb potential as well
as of the nucleon wave functions, break the factorized approach at the level of
cross-sections. Due to this,  most  
calculations of exclusive $(e,e'p)$ cross-sections are perfomed within a fully unfactorized
approach.
 Within the impulse approximation, the cross-section is  computed from the
 contraction of the electron current with the  nuclear current matrix element
of eq.~(\ref{nucc}). We recall now a relevant example.

\subsection{Matrix element of the electromagnetic current for free nucleons}
Let's assume for a moment that we compute the matrix element in  eq.~(\ref{nucc})
using free spinors  with momenta identical to the
asymptotic momenta of the nucleon (initial $\vec{p}_m$, final $\vec{P}_F$), 
so that the resulting nuclear matrix element
reads (omitting for the moment the spin indices):
\begin{equation}
J^{\mu}_{N,{\rm free}}(\omega,\vec{q},\vec{P}_f,\vec{p}_m)=
\delta(\vec{P}_F-\vec{p}_m-\vec{q})\bar{u}(\vec{P}_F)
 \hat{J}^\mu_N(\omega,\vec{q}\/) 
u(\vec{p}_m).
\label{free}
\end{equation}

In previous expression, 4-spinors and $4\times 4$ Dirac operators are involved. 
Besides, a  momentum-conserving delta function appears 
so that, in the computation of this matrix element,
 the momenta of the initial and final nucleon  are related by
$\vec{p}_m=\vec{P}_F-\vec{q}$.

\begin{figure}
  \begin{center}
    \parbox[b]{0.95\linewidth}{
      \includegraphics[width=0.85\linewidth]{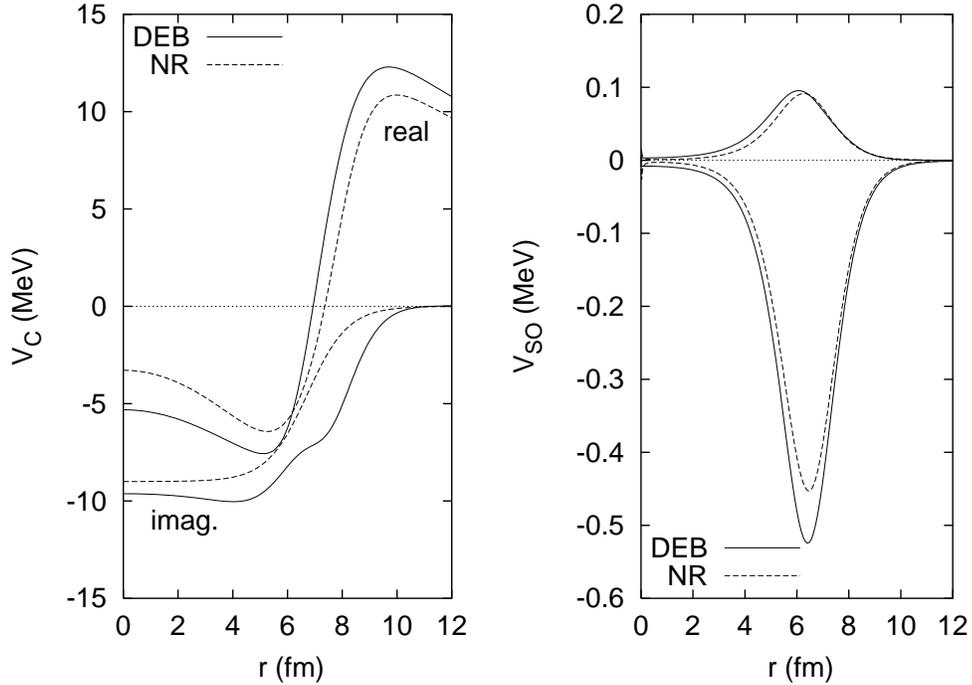}}
    \parbox[b]{7mm}{~}
    \parbox[b]{0.75\linewidth}{
      \caption[deb potentials]{\label{fig:deb}
     Comparison of standard nonrelativistic (NR) central 
and spin-orbit potentials for
$^{208}Pb$ at $T'=100$ MeV taken from ref.~\protect\cite{Quint}
and the potentials $V_C$, $V_{SO}$ derived with  the Dirac equation based (DEB)
procedure of eq.~(\protect\ref{debpot})
from a typical relativistic set of S-V potentials from ref.~\protect\cite{CHCM93}.}}
  \end{center}
\end{figure}

Now we note that
 the previous expression eq.~(\ref{free}) can be recast in a way that only Pauli spinors
appear explicitly:
\begin{equation}
J^{\mu}_{N,{\rm free}}(\omega,\vec{q},\vec{P}_F,\vec{p}_m)= 
\delta(\vec{P}_F-\vec{p}_m-\vec{q})
\chi^{\sigma_F\dagger} 
\hat{J}^{eff}_{2\times 2}({\rm free},\omega,\vec{q},\vec{P}_F,\vec{p}_m)
 \chi^{\sigma_i}
\label{nucc2}
\end{equation}

The effective $2\times 2$ current operator makes the bridge from the $4\times 4$
matrix expression to the one  suitable for a nonrelativistic
calculation. Explicit expressions for the nontruncated operator 
 $ \hat{J}^{eff}_{2\times 2}({\rm free}) $
can be found
in many references, for instance in~\cite{Ama96,Jeschonnek}. Otherwise, 
both expressions eq.~(\ref{nucc}) and
eq.~(\ref{nucc2}) are completely equivalent {\em for free spinors} in order
to compute the amplitude at the nucleon vertex.
For instance, the elementary electron-proton cross-section $\sigma_{ep}$ of eq.~(\ref{fac})
will typically be computed from the matrix element of eq.~(\ref{free}) or
equivalently eq.~(\ref{nucc2}). The $\sigma_{cc1}$ of de Forest 
is computed this way~\cite{Fore83}.

 If  
an operator
different from $\hat{J}^{eff}_{2\times 2}({\rm free},\omega,\vec{q},\vec{P}_F,\vec{p}_m)$ is employed
to compute the $(e,e'p)$ cross-section, 
factorization 
will obviously be lost, already at the level of the nuclear matrix element. 
 An example: in typical nonrelativistic calculations~\cite{DWEEPY},
   $\hat{J}^{eff}_{2\times 2}({\rm free},\omega,\vec{q},\vec{P}_F,\vec{p}_m)$
is expanded
  in powers
of $\vec{p}_m/M$ and as a consequence the
matrix element obtained with this truncated operator will yield numerically
different results from the one in eq.~(\ref{free}) or eq.~(\ref{nucc2}), {\em even when
sandwiched between free spinors}. This kind of difference between a nonrelativistically
truncated operator and $\hat{J}^{eff}_{2\times 2}({\rm free},\omega,\vec{q},\vec{P}_F,\vec{p}_m)$ 
is what we term {\em kinematical} (trivial) relativistic effects. These differences
can be avoided by using the full $\hat{J}^{eff}_{2\times 2}({\rm free})$.

\subsection{Further (nontrivial) differences between relativistic and nonrelativistic 
matrix elements}

According to the (relativistic) IA, to compute the nuclear matrix element
for the $(e,e'p)$ proccess, we substitute in eq.~(\ref{nucc}) the 
free spinors by  4-spinor solutions of  Dirac equation. Or alternatively,
in the (nonrelativistic) IA, we start from eq.~(\ref{nucc2}) and we substitute the free Pauli spinors
by solutions of the Schr\"odinger equation.
This introduces two kinds of numerical differences between approaches based on 
eq.~(\ref{nucc}) and eq.~(\ref{nucc2}) what we call {\em dynamical} differences:
\begin{enumerate}

\item The lower component of the 4-spinor solutions of Dirac equation
are not related to the upper ones by the same relationship (free Dirac equation)
that holds for free spinors and  that is implicitely assumed in the derivation of
$\hat{J}^{eff}_{2\times 2}({\rm free},\omega,\vec{q},\vec{P}_F,\vec{p}_m)$. 

\item The upper component of the solutions of Dirac equation are different from 
the solutions of Schr\"odinger one, even if equivalent potentials
(that lead to the same binding energy and phase-shifts) are used in both equations.
\end{enumerate}

We shall first focus on this second point. We usually start from the Dirac
equation:

\begin{equation}
(\tilde{E}\gamma_0-\vec{p}\cdot\vec{\gamma}-\tilde{M})\psi =0
\end{equation}
with S and V  spherically symmetric potentials,
\begin{eqnarray}
\tilde{E}&=&E-V(r) \\
\tilde{M}&=&M-S(r) \\
\psi&=&\left(\begin{array}
{@{\hspace{0pt}}c@{\hspace{0pt}}}
\psi_{up} \\
\psi_{down}\end{array}\right)\; ,
\end{eqnarray}             

This equation can be written either as a system of coupled linear differential equations
for $\psi_{up}$, $\psi_{down}$, or as a second order differential Schr\"odinger
like-equation for $\psi_{up}$: 

\begin{equation}
\left[\frac{-\vec{\nabla}^2}{2M}-U_{C}-U_{LS}\vec{\sigma}\cdot\vec{\ell}+
\frac{1}{2MrA(r)}\frac{dA(r)}{r}\vec{r}\cdot\vec{\nabla}\right]\psi_{up}(\vec{r})=
0
\end{equation}

Due to the linear term  $\vec{\nabla}$ (Darwin term), previous equation is not yet 
quite a Schr\"odinger-like equation.
Using the standard transformation
\begin{equation}
\psi_{up}(r)=K(r)\phi(r),
\label{darwin}
\end{equation}                    
the non-local (Darwin) term can be eliminated
to obtain a more standard Schr\"odinger equation with second derivatives only
\begin{equation}
\left[\frac{-\vec{\nabla}^2}{2M}-U_{DEB}\right]\phi(\vec{r})=
\frac{(E^2-M^2)}{2M}\phi(\vec{r})
\label{eqdeb}
\end{equation}

\begin{eqnarray}
U_{DEB}&=&V_C+V_{SO}\,\,\vec{\sigma}\cdot\vec{\ell} \nonumber \\
V_C &=&\frac{1}{2M}\left[V^2-2EV-S^2+2MS+V_D\right]\nonumber \\
V_D &=&\frac{1}{rA}\frac{\partial A}{\partial r}+\frac{1}{2A}
      \frac{\partial^2A}{\partial r^2}-\frac{3}{4A^2}
      \left(\frac{\partial A}{\partial r}\right)^2 \nonumber \\
V_{SO} &=&\frac{1}{2M}\frac{1}{rA}\frac{\partial A}{\partial r} \nonumber \\
A(r) &=&\frac{\tilde{E}+\tilde{M}}{E+M}=K^2(r) \; .
\label{debpot}
\end{eqnarray}

The former set of equations eqs.~(\ref{eqdeb}), (\ref{debpot}) 
defines the Schr\"odinger-like equation
and the Dirac equation based  (DEB) potentials. Eq.~(\ref{eqdeb})
is formally identical to the  usual Schr\"odinger equation, and actually
several groups use or have used  this analogy to compute
the full solution of the Dirac formalism taking advantage of computational tools
for nonrelativistic equations~\cite{Meucci,Kel97}.

A comparison of $V_C$ and $V_{SO}$ taken from the 
phenomenology with a nonrelativistic approach or with  the relativistic one after a DEB procedure,
 shows that
these potentials  are very similar, as depicted in Fig.~\ref{fig:deb}.
After careful inspection of the equations,
three remarks are however in order:

\begin{enumerate}
\item The solution of the Schr\"odinger-like equation, even though being  
phase-shift
and energy eigenvalue equivalent to the solution of the original Dirac equation,
is not identical to the upper component of the initial Dirac equation. It differs from
it by
the Darwin factor given by eqs.~(\ref{darwin},\ref{debpot}). 
\item The central potential $V_{C}$ (DEB) has a linear dependence on the energy.
That is, it is strongly nonlocal. Disregarding $V_{D}$ that 
is usually negligible, 
it is clear from previous eqs.~(\ref{debpot}) that, even if  V and S potentials of the
Dirac equation were energy independent,  $V_C$ (DEB) is not.
\item The lower component of the full 4-spinor solution of Dirac equation
is absent in the Schr\"odinger-like approach.
\end{enumerate}

We now study more in detail  points 1 and 2.

\subsection{Nonlocality and quenching of the upper components}

In the early times,   it was found in  analyses of nucleon-nucleus
reactions  based on the Schr\"odinger  equation  with
a  local optical potential, that this potential has a strong dependence on
the energy of the nucleon. This was attributed to the fact that
a fully nonlocal Schr\"odinger equation should be
used instead of the local one  given by eq.~(\ref{eqdeb}). The solutions of the
local and the nonlocal Schr\"odinger equation are related by  
a nonlocality factor \cite{Perey,Giani,Rawitscher},
in very much the same way  the Darwin factor relates the upper component of Dirac
equation to the solution of the DEB Schr\"odinger equation.

\begin{figure}
  \begin{center}
    \parbox[b]{0.95\linewidth}{
      \includegraphics[width=0.85\linewidth]{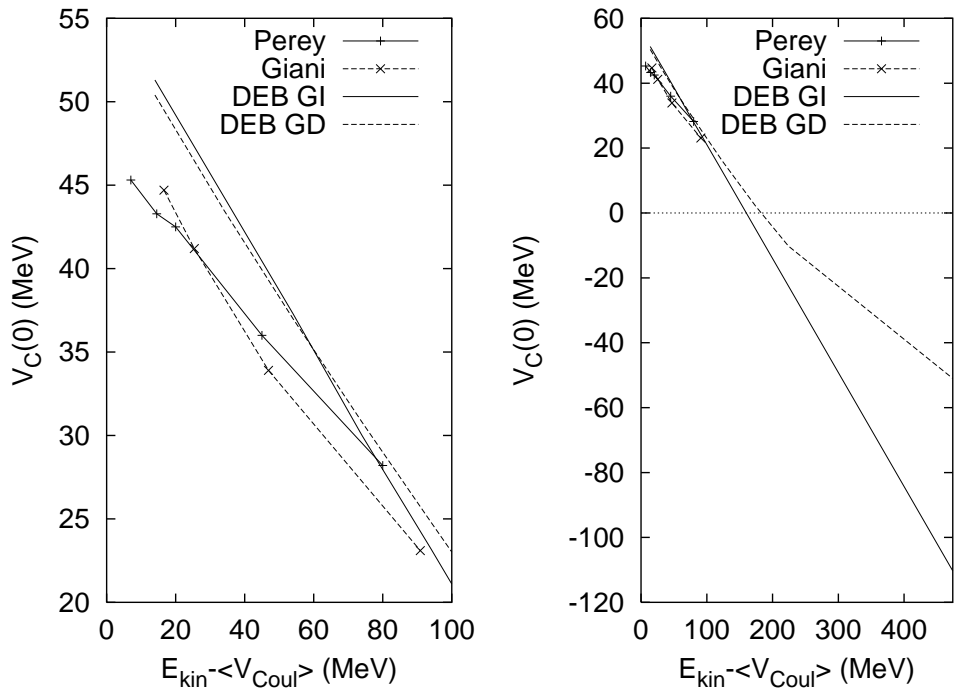}}
    \parbox[b]{7mm}{~}
    \parbox[b]{0.75\linewidth}{
      \caption[deb potentials]{\label{fig:nonlocal}
     Depth at the origin of the local potential  extracted from nonrelativistic analyses of
elastic nucleon scattering in $^{208}Pb$ at two different ranges  of kinetic energy of the scattered
nucleon.  Results from the well-known works of Perey {\em et al.}~\protect\cite{Perey}
and Gianini {\em et al.}~\protect\cite{Giani} are compared to the dependence found
in  $V_C$ (DEB)  taken from the S-V potentials of Clark {\em et al.}~\protect\cite{CHCM93}
at $T'=100$ MeV (DEB GI) and for the energy dependent relativistic parameterization (DEB GD).
}}
  \end{center}
\end{figure}

\begin{table}
\begin{center}
\begin{tabular}{|c|c|c|cc|}
\hline
   & particle & energy range & $V_{nl}^o$(MeV) & $ \beta_{nl}$ (fm)\\
\hline
Perey and Buck (1961)\cite{Perey} & n & 7-14.5 MeV & -71 & 0.85 \\
Giannini and Rico (1976)\cite{Giani} & n,p & 70-150 & -89 & 0.95 \\
Schwandt (1982)\cite{Schwandt} & p  & 80-150 & -75 & 0.82 \\
Clark GI & p  & 60 & -150 & 0.95 \\
Clark GD & p  & 22-1024 & -110 & 1.0 \\
\hline
\end{tabular}
\caption[Table I]{Nonlocality parameters of the Perey-Buck kind~\protect\cite{Perey}
 derived from optical potential analyses of
elastic nucleon-nucleus scattering. The depth $V^0_{nl}$ and nonlocality range
of the nonlocal (energy independent) potential equivalent to the local 
(energy dependent) one are quoted. Last two rows show what
we find if the same analysis is done to the $V_C$ potentials of the
DEB procedure from the S-V potentials of Clark et al. at $T'=100$ MeV (GI),
and for the EDAI parameterization (GD)~\protect\cite{CHCM93}. 
  \label{nonlocal.tab} }
\end{center}
\end{table}

In Fig.~\ref{fig:nonlocal} the energy dependence of several local optical potentials 
  is shown, along with the one 
of  $V_C$ (DEB). These local potentials
were fitted to   elastic nucleon scattering observables. In the case 
of the DEB ones, they were obtained from the S-V  potentials
of ref.~\cite{CHCM93}, in one case neglecting the additional energy dependence of
the relativistic potentials,  that is, introducing in eq.~(\ref{debpot}) 
the S and V potentials  for a fixed kinetic
energy
of  100 MeV (DEB GI), and in the other case (DEB GD) the  rather soft dependence of
the S, V potentials is included, thus in eq.~(\ref{debpot}), 
besides from the explicit linear dependence on E, we also have S(E) and V(E),
the energy dependence of the S and V potentials. This additional energy dependence
is  only visible if a large range of kinetic energy is chosen, as done
 in the right panel of Fig.~(\ref{fig:nonlocal}). On the other hand, as shown
in the left panel, for a small  range of kinetic energies, the dominant dependence on the
energy of the DEB potential is just the linear one explicitly seen in eq.~(\ref{debpot}).

 \begin{figure}
       \begin{center}
         \parbox[b]{0.95\linewidth}{
           \includegraphics[width=0.85\linewidth]{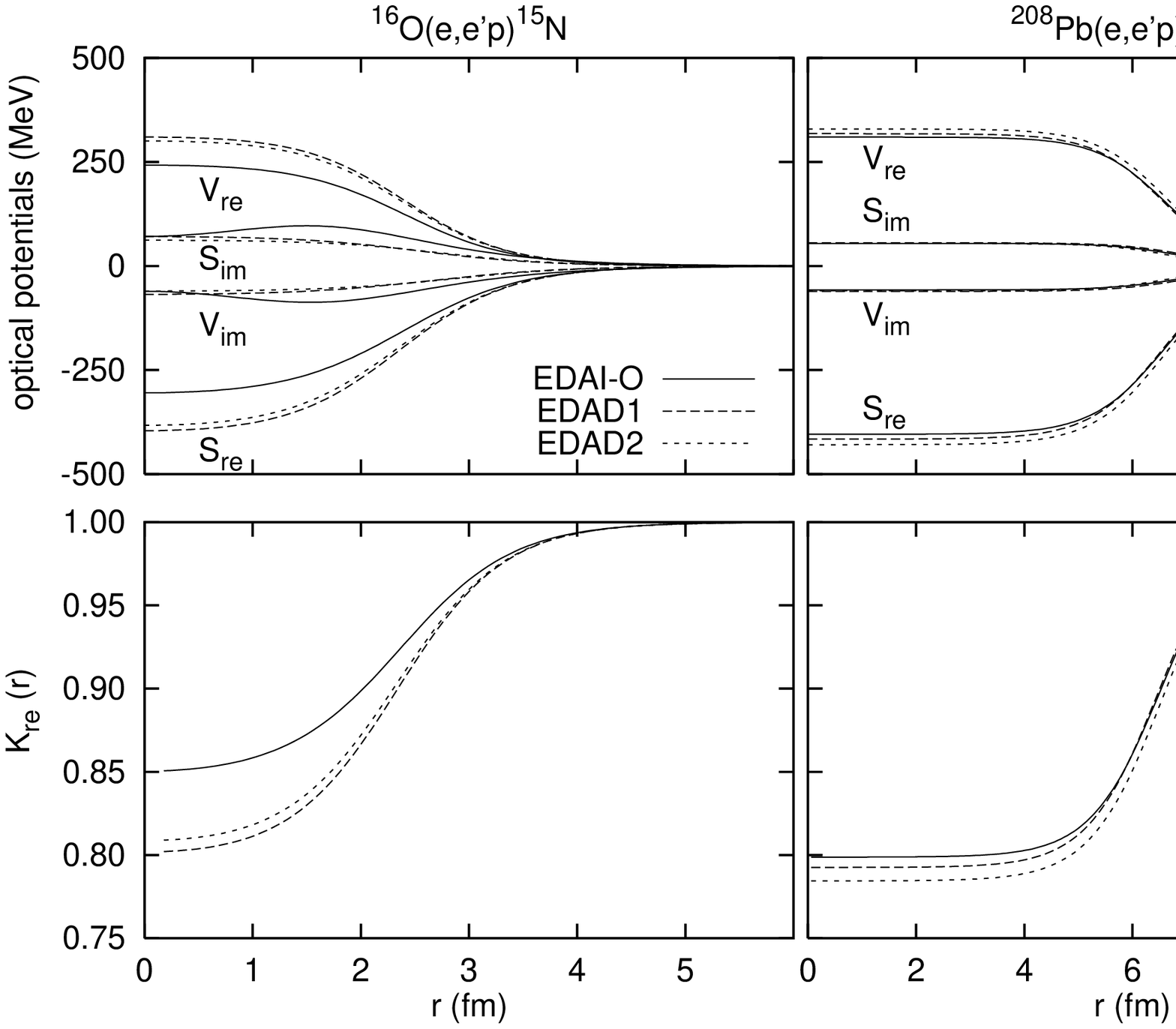}}
       \parbox[b]{7mm}{~}
       \parbox[b]{0.75\linewidth}{
         \caption[deb potentials]{\label{fig:optpot.rel.ps}
Upper pannels: relativistic S-V optical potentiasl for $^{15}N$ (left) and $^{207}Tl$ (right) at
$T'=100$ MeV, from three different parameterizations. Lower pannels:  Darwin
factors associated to the relativistic optical potentials. Only the real part of the Darwin
term is shown, the imaginary one is negligible.
We see that, while for $^{207}Tl$  the A-dependent (EDAD-1 or EDAD-2) and A-independent parametrization 
of the
potentials 
lead to very similar results. However, for $^{15}N$
there are noticeable differences both in the potentials and the
Darwin term. These differences translate in a large variation of the scale factor derived from
the A-independent and the A-dependent potentials for the reaction $^{16}O(e,e'p)^{15}N$ 
(see Fig.~\protect\ref{fig:espect}).
   }}
     \end{center}
   \end{figure}

Also we quote in Table~\ref{nonlocal.tab} the parameters of the nonlocal
potential (but energy independent) that leads to same elastic proton scattering observables than the
local one (but energy dependent), as well as the ones that result when a
nonlocality analyses of the Perey kind~\cite{Perey} is done to the DEB potentials. The
nonlocality range of the potential is similar in all cases, ($\beta_{nl}$)  in between 0.82 and 1.0 fm.

A larger
value  of the nonlocality range implies a  Perey
factor more different from one. The shape of the Perey factor and that of the Darwin
factor is very similar (for instance the Darwin factor can be seen in Fig.~\ref{fig:optpot.rel.ps})
More remakably, if one considers a DEB potential and builds  
the Perey factor from the  linear energy dependence displayed
by the potential,  then this factor
is practically equal to the Darwin factor needed to make the solution of
the Schr\"odinger-like equation $\phi$  in eq.~(\ref{eqdeb}) equal to the upper component
 of the solution
of the initial Dirac equation~\cite{Udi95}.
This very striking equivalence  has been 
 studied in high detail in~\cite{Rawitscher,Udi95,Cannata}. 

Thus, the situation is such that the nonlocal nature of the optical potential
seen in the nonrelativistic analyses could very well be a way of implementing the
relativistic features of Dirac equation. But also perhaps  
the other way around!. Only
with elastic observables it is not possible to disentangle between these two
(nonlocality of the Perey kind  {\em versus} Darwin factor from Dirac equation) aspects, as they
are not sensitive to the inner part of the wave function. $(e,e'p)$ experiments
on the other hand, do sample the nuclear interior where either the Darwin or the Perey factor 
reduce the nucleon density. This  yields  smaller  cross-section and thus
has  an effect in scale factors~\cite{Udi93,Udi95}.
However,  given this equivalence between Perey and Darwin factors (at least for intermediate
nucleon energies) both relativistic and nonrelativistic approaches would be 
also impossible to disentangle attending only to cross-sections or
scale factors derived from $(e,e'p)$ experiments. As we will see later,
 perhaps other observables different from cross-sections will be of help.

A few additional comments: a) The Darwin or Perey term controversy could in principle be
also present  for the
bound state wave function. However, the shape of the Darwin term is almost
flat (see Fig.~\ref{fig:optpot.rel.ps}) except at the nuclear surface, where it goes
to one.  Taking into account that the bound state is normalized to one in the nuclear
volume, the effect of an almost
constant factor inside  this nuclear volume, as it is the case  of the Perey or Darwin ones, 
is negligible.
 Recall that, on the other
hand, the scattering states are normalized by comparing their asymptotical behavior with 
the one of 
free waves, that is, in a region where both the Perey and the Darwin factor are equal to
one and thus the normalization procedure does not affect them.
b) Even though the Perey kind of nonlocality associated to a  linear dependence in the energy of
the potentials is
the dominant one, it is possible that additional nonlocalities 
should be considered. We refer the reader to ref.~\cite{Rawitscher} for a detailed account
of the comparison of extended nonlocality nonrelativistic approaches and the relativistic
one.

\subsection{Enhancement of the lower components}

So far, there is one important difference between relativistic and nonrelativistic 
approaches that remains to be studied, namely the lower components of the Dirac equation solution. 
Nonrelativistic formalisms are based on the effective $2\times 2$ operator
 $\hat{J}^{eff}_{2\times 2}({\rm free},\omega,\vec{q},\vec{P}_F,\vec{p}_m)$ derived after making
some sort of assumption about the lower components, most often, that they are given by the same relation
that holds for free spinors. This assumption is also made in computing $\sigma_{ep}$ of eq.~(\ref{fac}),
and thus it is a prerequisite for a factorized calculation. For a general solution of Dirac equation
we have:

\begin{equation}
\psi_{down}=\frac{\vec{\sigma}\cdot\vec{p}}{\tilde{E}+\tilde{M}}\psi_{up},   
\label{lower}
\end{equation}
for a free nucleon $\tilde{M}=M$, $\tilde{E}=E$, but in a Dirac equation with typical standard
 potentials,
we have $\tilde{E}+\tilde{M}\simeq 0.5 (E+M)$
thus the lower components are {\em enhanced} with respect to the free case.
Due to this, matrix elements of eq.~(\ref{nucc}) have a different behaviour than
the ones of eq.~(\ref{free}) and thus the {\em matrix element}   does not factorize into
a {\em free part} dealing only with free spinors (positive energy solutions) and 
the nucleon momentum distribution (see for details refs.~\cite{Cab98a,Cab98b}).
Thus factorization is broken even before summing and averaging in spins  and
integrating into the bound nucleon momentum to obtain the matrix element.
This is quite different from the nonrelativistic case where there is always factorization
at the level of the matrix element, that is, before the integration in $\vec{p}$ of
the bound nucleon and  summing and averaging of spins, provided that the
same operator $\hat{J}^{eff}_{2\times 2}({\rm free},\omega,\vec{q},\vec{P}_F,\vec{p}_m)$ 
is used to compute $\sigma(e,e'p)$ and $\sigma_{ep}$. As it is well known \cite{BGPR96}
in the nonrelativistic IA, 
the spin-orbit part of the optical potential for instance  breaks factorization at the level
of the cross-section (after summing and averaging over spins). Thus if no FSI effects
are considered (using plane waves  for the final nucleon), factorization may be
recovered in the nonrelativistic IA. However, as we have just mentioned,
factorization is broken in the
relativistic case even if 
no FSI are included in the calculation~\cite{Cab98a}. This {\em strong} breakdown of
factorization in the relativistic case is due to the enhancement of the lower components
with regards to the free case and should translate into differences in the $(e,e'p)$ observables.
In particular  the relativistically computed  cross-sections should depart {\em more strongly}
 from the
factorized result than the nonrelativistic one.

\begin{figure}
  \begin{center}
    \parbox[b]{0.95\linewidth}{
      \includegraphics[width=0.75\linewidth]{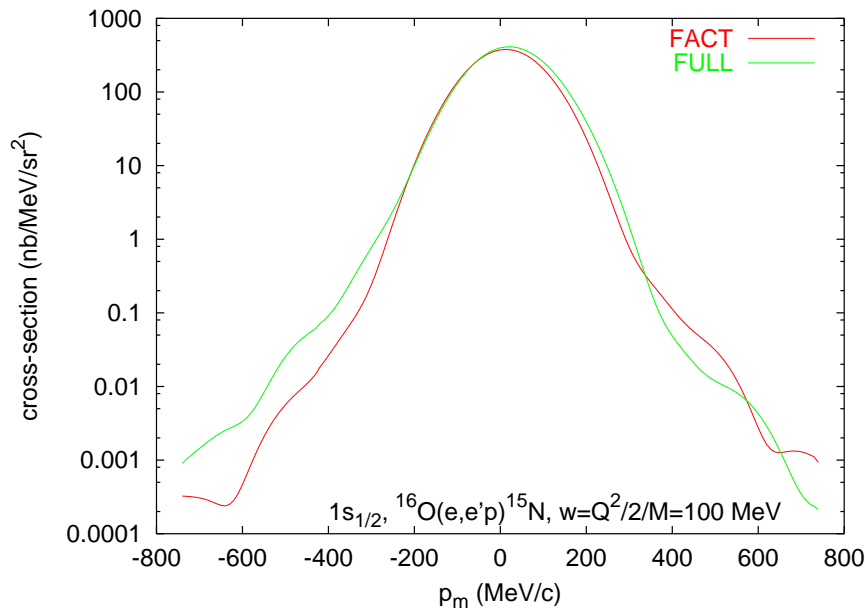}}
    \parbox[b]{7mm}{~}
    \parbox[b]{0.75\linewidth}{
      \caption[espect]{\label{fig:cross}
Cross-section for the  $^{16}O(e,e'p)^{15}N$ reaction. The  
full
calculation is compared with the  factorized result.
 }}
  \end{center}
\end{figure}

\begin{figure}
  \begin{center}
    \parbox[b]{0.95\linewidth}{
      \includegraphics[width=0.85\linewidth]{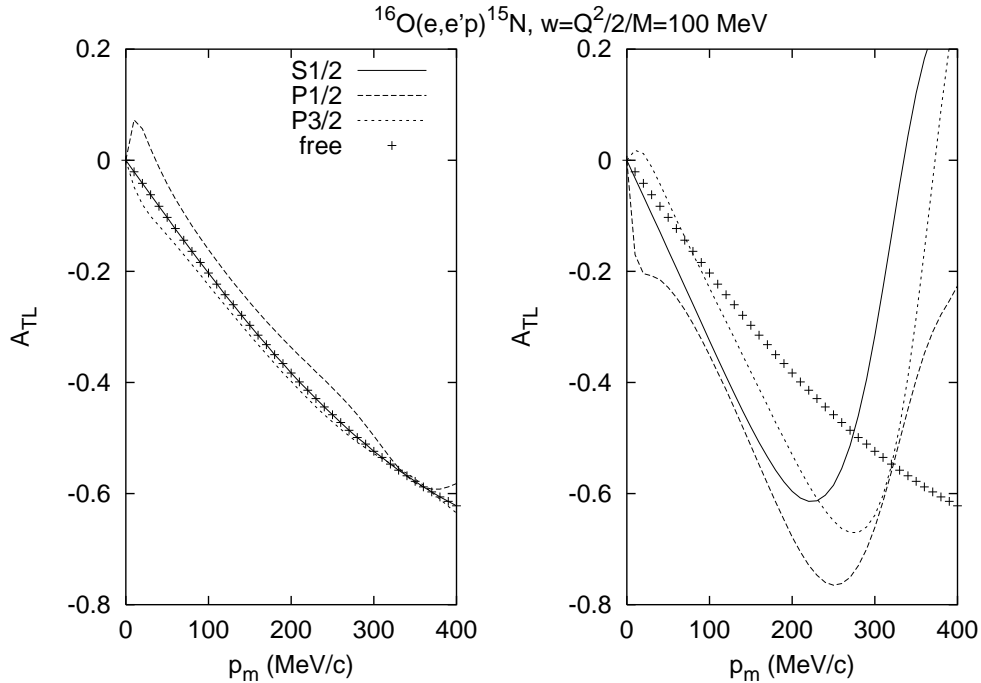}}
    \parbox[b]{7mm}{~}
    \parbox[b]{0.75\linewidth}{
      \caption[espect]{\label{fig:atl}
Effect of the dynamical enhancement of the lower components observed in
 $A_{TL}$.
In the right panel the dynamical enhancement as well as the  
dispersive distorsion
of the spinors is removed. The crosses  show the free result, identical
to the   $s-$wave result in this approximation. In the left panel, the
enhancement of the lower components and the dispersive distortion of the
spinors is switched on. The departure from the factorized (free) result and
the richer structure for the three shells is clearly seen. Note the  tendency of
the $j=\ell-1/2$ spin-orbit partner to depart more from the free result\protect\cite{Udi99,Cab98b}. 
 }}
  \end{center}
\end{figure}

As shown in Fig.~\ref{fig:cross}
factorization is  a good estimate of the `bulk'  of the cross-section, at least for quasielastic
kinematics. A breakdown of the factorization would be easier to see 
measuring partial contributions to the cross-section, and comparing them  to the $L$, $T$ and $TL$
contributions to $\sigma_{ep}$.
Factorization  means that the reduced spectral function
$\rho$:

\begin{equation}
\rho=\frac{d^5\sigma(e,e'p)}{d\Omega_e d\Omega_F d\varepsilon'}
(K \sigma_{ep})^{-1}
\end{equation}
would be the identical if derived from  the total contribution to both 
the $(e,e'p)$ cross-section in the numerator and
the electron-proton one in the denominator, or only the $T$, $L$,
$TL$ or any other individual contribution to the cross-section.
Now, instead of comparing  partial contributions to the reduced cross-sections 
(or the response functions $R^{L}, R^{T}, R^{TL}$), 
it will be better to compare
ratios of cross-sections, that are independent on the scale-factor, and thus would
clearly display any departure from the factorized result. In particular
the TL asymmetry$A_{TL}$ given by
\begin{equation}
A_{TL}=
\frac{\sigma(\phi_F=180^o)-\sigma(\phi_F=0^o)}
{\sigma(\phi_F=180^o)+\sigma(\phi_F=0^o)} \; .
\end{equation}                  
is very adequate.
In Fig.~\ref{fig:atl} this observable is displayed for the $1s_{1/2}$, $1p_{3/2}$ and $1p_{1/2}$ shells of
$^{16}O$. If factorization
were employed in the calculation, then  the $A_{TL}$ curves for the three
shells  will be identical to the free result. In the left panel
we show results obtained when the enhancement of the lower components  as well
as the dispersive distortions of the spinor are removed. For these results
factorization is only broken due to the spin-orbit coupling in the final
state. In particular for the $s-$wave, factorization is fully recovered
(factorization of the cross-section
can be recovered if  there is no spin-orbit in the final state and/or
if the initial state is an $s-$ state. This  is due to the particular angular
momentum algebra of the spin 1/2  final particle~\cite{Jathesis}).
For the other two shells, only a modest deviation from the factorization
approach is seen. This is typical for quasielastic kinematics.

On the right panel, however, the whole calculation 
including the enhancement of the lower components is displayed. This shows
a very visible departure from the factorized result for $A_{TL}$, with a 
much richer structure and a crossover of $A_{TL}$ in between 200 and 300
MeV. This is a  significant feature of $A_{TL}$ that has been 
 confirmed by  experiment~\cite{Gao2000}.
Also, the departure is larger for the $j=\ell-1/2$
spin-orbit partner as predicted by the relativistic IA result~\cite{Cab98b,Udi99}.

\begin{figure}
  \begin{center}
    \parbox[b]{0.95\linewidth}{
      \includegraphics[width=0.85\linewidth]{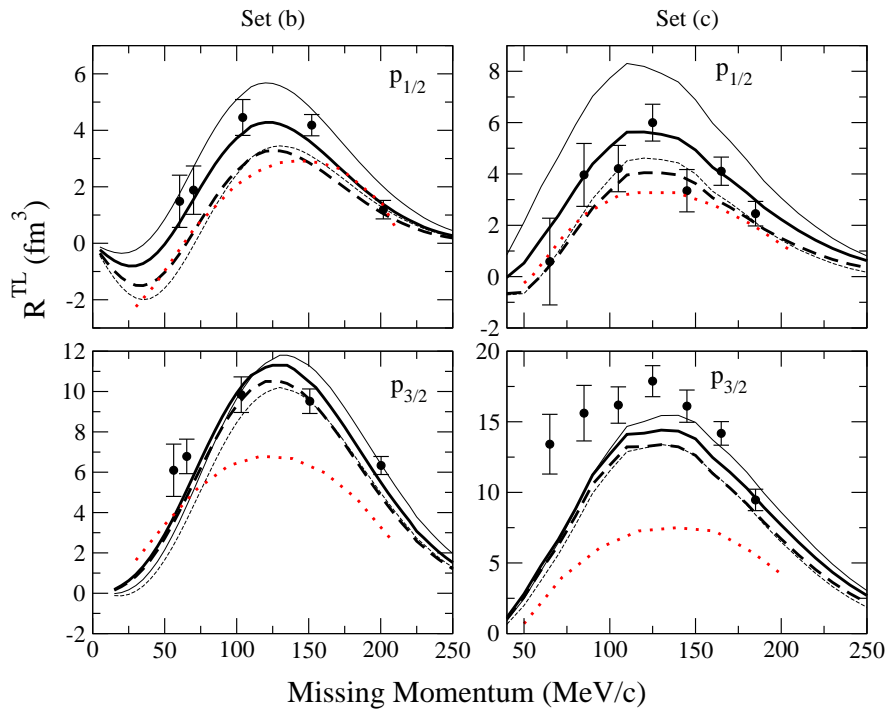}}
    \parbox[b]{7mm}{~}
    \parbox[b]{0.75\linewidth}{
      \caption[espect]{\label{fig:rtl}
$TL$ responses compared to experimental data  for $^{16}O(e,e'p)^{15}N$.
Experiments from
Spaltro et al.~\protect\cite{Spal93} (set (c), $Q^2$ of around 0.2 $(GeV/c)^2$), and
Chinitz et al.~\protect\cite{Chin91} (set (b), $Q^2$ of around 0.3 $(GeV/c)^2$)
Results with the current operator $cc1$ (thin lines) and $cc2$ (thick lines), 
 Coulomb gauge, NLSH-P
wave functions\protect\cite{Udi2001} and EDAI-O optical potential. Solid lines show the results
of the full calculation, dashed lines the results without the dynamical enhancement of the
lower components. Dotted lines show the result of the nonrelativistic analyses of 
ref.~\protect\cite{Spal93}.
 }}
  \end{center}
\end{figure}

The important effect of the enhancement of the lower components in the $A_{TL}$ observable
would also be seen in the separate response $R^{TL}$ and indeed this has been 
confirmed by experiments at $Q^2=0.8$ $(GeV/c)^2$\cite{Gao2000}.
We have then also compared the relativistic IA formalism to former  results for $R^{TL}$
obtained  at moderate values of
$Q^2$ (see Fig.~\ref{fig:rtl}) obtained at Saclay~\cite{Chin91} and  NIKHEF~\cite{Spal93}.
Though the  enhancement of the lower components (present in the solid curves and removed in the
dashed ones)  does not show up as clearly  in $R^{TL}$ at these low $Q^2$ kinematics,
the overall agreement of the relativistic IA result is better than that of the
 nonrelativistic analyses~\cite{Spal93} (dotted curve in Fig~\ref{fig:rtl}). 
 
\section{Scale factor analyses over a wide range of $Q^2$}
   
\begin{figure}
  \begin{center}
    \parbox[b]{0.95\linewidth}{
      \includegraphics[width=0.85\linewidth]{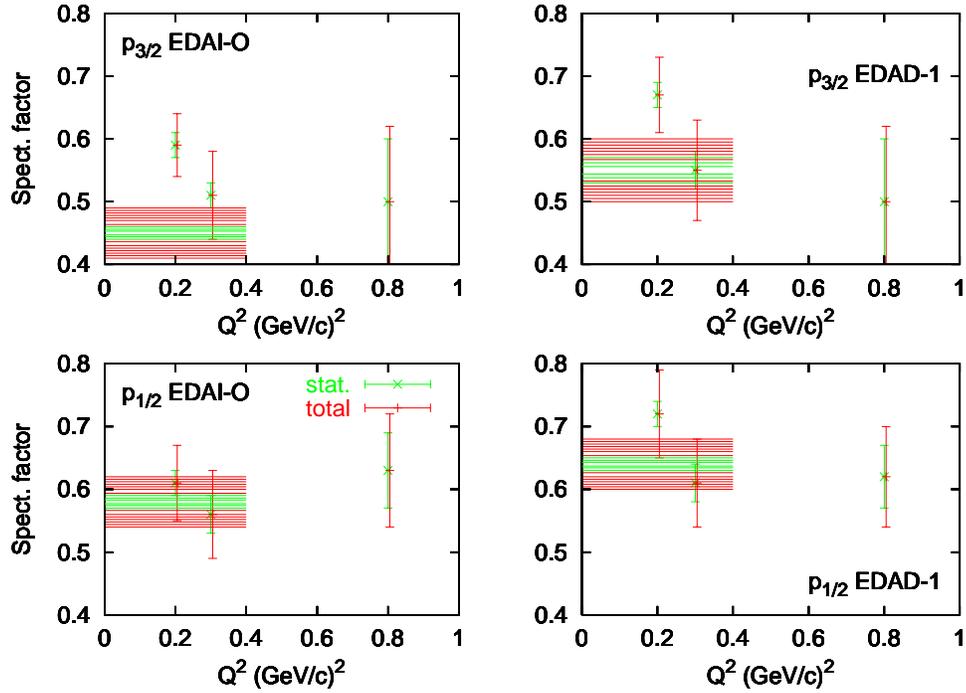}}
    \parbox[b]{7mm}{~}
    \parbox[b]{0.75\linewidth}{
      \caption[espect]{\label{fig:espect}
Scale factors derived from experimental data for the valence states in $^{16}O(e,e'p)^{15}N$ 
at several $Q^2$. Experiments from
Leuschner et al.~\protect\cite{Leus94} ($Q^2$ range from 0 to 0.4 $(GeV/c)^2$, 
represented by the dashed area),
Spaltro et al.~\protect\cite{Spal93} ($Q^2$ of around 0.2 $(GeV/c)^2$),
Chinitz et al.~\protect\cite{Chin91} ($Q^2$ of around 0.3 $(GeV/c)^2$)
and Gao et al.~\protect\cite{Gao2000} ($Q^2$ of around 0.8 $(GeV/c)^2$) are
shown. Results with the current operator $cc2$,  Coulomb gauge, NLSH-P
wave functions and EDAI-O (left) and EDAD-1 (rigth) potentials~\protect\cite{Udi2001}.
 Statistical only
and total uncertainty error bars are shown. Note that due to fragmentation, part of
the $3/2^+$ strength is not included in the lowermost state here analyzed, what causes 
the scale factors for the $p_{3/2}$ shell to be smaller than for the $p_{1/2}$ one.
 }}
  \end{center}
\end{figure}

Some recent works~\cite{Zhalov,Lapikas} point to the possibility of a $Q^2$
dependence of the spectroscopic factor.  For $^{16}O$ there are several
$(e,e'p)$ experiments performed at different value of $Q^2$, so it is
a good nuclues to look for such effect. In Fig.~\ref{fig:espect}, the scale factors needed to
scale the theoretical RDWIA reduced cross-section to the experimental data are compared.
Four sets of data points were used, one in parallel kinemtatics from Leuschner {\em et al.}
\cite{Leus94}, and three in perpendicular kinematics (from refs.~\cite{Spal93,Chin91,Gao2000}).

We must note the following:
\begin{enumerate}
 \item Scale factors are not independent on final state interactions (FSI). In particular, 
calculations with FSI based upon optical
potentials fitted to elastic proton scattering in one hand, 
and upon the Glauber approach on the other, give
generally different scale factors\cite{Jan,Lapikas}. This is due to the different
picture of the FSI interaction assumed in each case. Which one is dominant in $(e,e'p)$
processes is yet to be known, and probably depend on the kinematics.
 
 Also, as already said here, elastic proton data mainly constrain the  
asymptotic behaviour of the optical potentials.  $(e,e'p)$
experiments are not very sensitive to this asymptotic or large r region, 
rather to the behaviour of the proton wave function (and thus the potential) in the inner nuclear
region. This is the reason  why optical potentials that yield essentially the same
elastic $(p,p)$ observables, can however lead to $(e,e'p)$ scale factors that
differ by 50\% or more~\cite{Udi2001}. In  Fig.~\ref{fig:espect}, the results
shown in the left and
right panels differ only in the relativistic optical potential employed. 
The
effect on the scale factor is clearly visible while none of the
two can be  preferred over the other based upon elastic $(p,p)$ scattering only. 
A very effective way of dealing with this optical potentical uncertainty is
to use also data from {\em inelastic} nucleon scattering, restricting thus
more the potentials in the nuclear interior~\cite{KellyEEI}. This is at the moment only
available for nonrelativistic potentials  in a restricted range of energies.

For a heavier nucleus as  $^{208}Pb$ scale factors were much more stable against different choice 
of the optical potentials~\cite{Udi93},
 mainly because all the available relativistic parametrizations were
much more similar than for $^{16}O$ (compare left and right panel of Fig.~\ref{fig:optpot.rel.ps}).

\item Usually, lower $Q^2$ experiments span a smaller range in $p_m$ than large $Q^2$ ones.
Due to this, large $Q^2$ experiments  give more weight to the large $p_m$
region, where the IA is not the only prevailing reaction mechanism. Other
contributions would naturally lead to an increase of the cross-section in
the $p_m$ region where the single-particle spectral function is not dominant,
and thus fitting scale factors with  data in this region
will yield higher scale factors. When comparing  data from 
experiments taken at different $Q^2$ in order to {\em deduce scale factors}, one should probably
restrict oneself to those data in region of $p_m$ where the
IA result   is important. So far this region is not well explored at high $Q^2$ for
$^{16}O(e,e'p)^{15}N$, but new experiments at Jlab aiming at this will yield new high quality data
very soon.

\end{enumerate}

\section{Summary}

In this contribution we have studied the differences between 
impulse  approximation calculations for the $(e,e'p)$ reaction based upon the Schr\"odinger  and Dirac equation.
The overall good agreement of the Dirac equation based results with experiment
 calls for a closer
reexamination of the relativistic approach. While it is true that relativity should play a role
in high $Q^2$ experiments, and thus a better agreement of the relativistic approach is expected
from kinematical grounds, the fact that dynamical predictions of the single-particle 
relativistic picture, as for instance  the richer structure of the $A_{TL}$ asymmetry and the 
enhancement of the $R^{TL}$ response are   confirmed by the experiment
needs further research where some approximations  of the model  (in particular its single-particle nature) 
could be removed.

\section*{Acknowledgments}

This work was partially supported under Contracts  No.
PB/98-1111, PB/98-0676, PB/96-0604
and by the Junta de Andaluc\'{\i}a (Spain). J.R.V. and A.E. acknowledge
support from  doctoral fellowships of the Consejer\'{\i}a de Educaci\'on
 de la Comunidad de Madrid and Ministerio de Educaci\'on y Cultura
 (Spain), respectively.

\end{document}